\documentstyle[aps,prl,eqsecnum,twocolumn]{revtex}
\begin{document}
\draft
\twocolumn[\hsize\textwidth\columnwidth\hsize\csname @twocolumnfalse\endcsname
\title{Do electrons change their $c$-axis kinetic energy upon entering the superconducting state?}
\author{Sudip Chakravarty}
\address{Department of Physics and Astronomy\\University of California Los
Angeles\\Los Angeles, California 90095-1547}
\date{\today}
\maketitle
\begin{abstract}
The interlayer tunneling mechanism of the cuprate high temperature superconductors involves a
conversion of the confinement kinetic energy of the electrons perpendicular to the CuO-planes ($c$-axis)
in the normal state to the pair binding energy in the superconducting state. This mechanism is discussed
and the arguments are presented from the point of view of general principles. It is shown that 
recent measurements of the $c$-axis properties support the idea that the electrons  substantially lower their
$c$-axis kinetic energy upon entering the superconducting state, a change that is nearly impossible in any
conventional mechanism. The proper use of a $c$-axis conductivity sum rule is shown to
resolve puzzles  involving the  penetration depth and the optical measurements.
\vskip 2.0cm
\end{abstract}
]
\section{Introduction}
There are abundant indications of a remarkable
mechanism of superconductivity in cuprate high temperature superconductors, known 
as the interlayer tunneling mechanism\cite{Anderson1}. At its simplest, the theory is that the confinement
kinetic energy of the electrons in the normal state
is converted into the superconducting binding energy. From the
uncertainty principle, confinement  implies  a  kinetic energy of order
$\hbar^2/2m d^2$, where $d$ is the separation between the planes. It is as though the electrons were confined in a
deep potential well perpendicular ($c$-axis) to the CuO-planes\cite{foot1}.  There is
considerable experimental support for this  idea, although
theoretical controversy persists. 

The very   concept
of confinement of the motion of the electrons is at
odds with  the time honored notion of a Fermi liquid. At issue is the concept of orthogonality
catastrophe in a non-Fermi liquid, which posits that the motion along the $c$-axis is accompanied by
the overlap of many  particle wave functions of $N$ electrons, vanishing as $N\to
\infty$.  In the absence of controlled non-perturbative methods to treat this inherently
non-perturbative phenomenon, little has been settled. It is therefore necessary to present the arguments
from the point of view of general principles. 

Consider the question posed in the title of this paper: ``Do electrons change their $c$-axis kinetic energy
upon entering the superconducting state?" It is useful to expand on the precise meaning of this question. In a
BCS superconductor, the kinetic energy of the superconducting state is {\em greater} than that of the
normal state. The reason is that the normal state is a Fermi liquid  in which the kinetic energy is
diagonal, the happiest possible situation from the point of view of the kinetic energy. Therefore, any change of
state must necessarily increase the kinetic energy. This increase is, however,  overwhelmed by the gain in the
potential energy. Thus, a BCS superconductor becomes a superconductor despite the increase in the kinetic energy.  In
contrast, if we considered the transition to a superconducting state from  a state in which the kinetic energy is
not diagonal, the driving mechanism can be the saving in the kinetic energy. The interlayer mechanism capitalizes on
the possibility that the $c$-axis kinetic energy is frustrated in a non-Fermi liquid. The question we ask is whether
or not  this frustration is relieved in the superconducting state, and whether or not the phenomenology of
the cuprates support this theory.

How should we view the  crossover from two to three dimensions in cuprate superconductors? In particular, what is
the role of the fluctuations of the phase of the superconducting order parameter? It will be shown that the
issues involving phase fluctuations are separate from the issues involving the microscopic
superfluid stiffness.  A striking characteristic of the
interlayer mechanism is that the coupling between the layers can significantly enhance this stiffness, which is
nearly impossible in a conventional BCS superconductor. The phase fluctuations should, however, be  similar to
those in a conventional superconductor. The prospect of unifying the concepts of phase fluctuations with
the concepts of interlayer tunneling then becomes apparent. 

How can we test the change of the $c$-axis kinetic energy? This question is answered  using  a
powerful sum rule for the
$c$-axis conductivity, which, it will be shown, leads to the resolution of an apparent paradox posed
by the optical measurements in these materials over the years. The paradox has been that the $c$-axis
penetration depth estimated from the change in the kinetic energy alone is apparently the same as that
obtained ignoring this change.

\section{Superconductivity as a 2$D$ to 3$D$ crossover phenomenon}
Superconductivity in cuprates  can be viewed as a  dimensional crossover between two (2$D$) and
three dimensions (3$D$). This is an experimental fact. While the charge
transport  perpendicular to the CuO-planes in the normal state is indicative of an insulator, there is
perfect coherence in the superconducting state. 
Dimensional crossovers are known even for classical
statistical mechanical problems involving phase transitions, or for quantum statistical mechanical
problems that can be effectively viewed in terms of order parameters with only classical fluctuations
at finite temperatures. What, then, is different here? To answer this question, it is necessary to probe
it more carefully.
  
Phase transitions  in classical statistical mechanics are
independent of the kinetic energy; only the potential energy is relevant. The superconducting transition in BCS
superconductors can be described by a classical complex order parameter
theory, namely the Ginzburg-Landau theory; quantum mechanics determines merely the parameters of this
model. Thus, the kinetic energy cannot play an explicit role in this phase transition. Low
dimensional superconductors are known to exhibit considerable fluctuation effects at finite
temperatures that are entirely classical in nature.  In dimensions less than or equal to two, the
fluctuations are so severe that the order parameter vanishes. In two dimensions, a topological phase
transition to a superconducting state takes place, but with a vanishing order paramete\cite{Kosterlitz}. In the low
temperature state, there is a finite superfluid density as determined from the current response, but no long range
order. 

Imagine now that two dimensional planes are stacked to form a three dimensional superconductor.
Conventionally, this is described by the Lawrence-Doniach (LD) model\cite{LD}, which consists of the
free energy functional
\begin{eqnarray}
{\cal F}&=&\sum_n\int d^2x \bigg[\alpha |\psi_n|^2+{1\over 2}\beta|\psi_n|^4\nonumber\\
&+&{\hbar^2\over 2m_{\rm
ab}}\left|\nabla\psi_n\right|^2+{\hbar^2\over
2m_cd^2}\left|\psi_n-\psi_{n+1}\right|^2\bigg]\label{LDF},
\end{eqnarray}
where $\alpha$, $\beta$, $m_{ab}$, and $m_c$ are parameters that are in general temperature dependent. 
The order parameter in the  plane labeled $n$, $\psi_n(x,y)$, is a function of  the $2D$ coordinates $x$
and $y$. The bending energy in the
$ab$-plane is expressed in terms of a gradient energy, but the energy in the perpendicular direction is
written in its discrete form. This is correct, because although the coherence length in the planes is
frequently much larger than the lattice spacing, it is not so in the perpendicular direction, and
therefore the  continuum limit cannot be taken in this direction. 
The minimization of this functional determines the order parameter in mean field theory, but to
incorporate  fluctuations it is necessary to integrate over all possible order parameter
configurations in the partition function.

This is emphatically a classical
model\cite{foot2}. What determines the coupling between the layers? It is argued that this is due to
the Josephson effect\cite{Tinkham}. Assume for the moment that the magnitude of the order parameter is
independent of $n$,
$\psi_n=|\psi|e^{i\phi_n}$. Then, the last term in Eq.~(\ref{LDF}) is
\begin{equation}
{\hbar^2|\psi|^2\over 2m_cd^2}\left[1-\cos(\phi_n-\phi_{n+1})\right]\ge 0.
\label{cosphi}
\end{equation}
This coupling can represent the Josephson effect only close to $T_c$, where the Josephson
coupling energy is indeed proportional to the square of the order parameter in a conventional
superconductor, while it is only proportional to the magnitude of the order parameter as $T\to
0$\cite{Chak1}. This is not terribly disturbing because the Ginzburg-Landau functional is only
supposed to be valid close to $T_c$. But it must be remembered that
there is {\em no} LD model at low temperatures for conventional superconductors, which is
a frequently misunderstood point\cite{Leggett}. In contrast, the Josephson effect between two
superconductors with non-Fermi liquid normal states can be recast in the language
of the LD model\cite{Chak1}. 

In mean field theory, the free energy
functional is minimized by setting the order parameter to be the same everywhere, that is, both its
magnitude and phase. If we apply this theory to Eq.~(\ref{LDF}), we are back to uncoupled  layers and no
enhancement of the mean field  transition temperature, $T_c^0$. Sometimes, an enhancement is claimed, which
is merely the result of considering the functional in Eq.~(\ref{LDF}) in which the coupling between the
layers is taken to be
$-(\psi_n^*\psi_{n+1}+ {\rm c. c.})$ instead of $|\psi_n-\psi_{n+1}|^2$. For a conventional superconductor
this is incorrect, because the Josephson energy in that case is proportional to
$[1-\cos(\phi_n-\phi_{n+1})]$ instead of $-\cos(\phi_n-\phi_{n+1})$. The only way to enhance $T_c^0$ would
be to change the parameters of the LD model appropriately. This is difficult to achieve in a BCS
superconductor, because the density of states is changed very little by the small hopping matrix elements
of the electrons between the layers of a highly anisotropic superconductor. 

So, what does the coupling between the layers do? It can suppress phase fluctuations by coupling 
the phases of the layers to raise the true $T_c$ closer to $T_c^0$.   Note that, in general,
the true $T_c$ is less than $T_c^0$.
 Thus, a dimensional crossover is
driven by suppressing phase fluctuations,  and unless the  individual two dimensional layers have a  
high $T_c^0$, we gain little by suppressing phase fluctuations. Of course,  phase
coherence will be established in all three directions, and those properties that depend on this
coherence will certainly be affected. The actual increase of the true transition temperature due to 
interlayer coupling can be easily estimated from the $XY$-model\cite{Kosterlitz}.

A prominent feature  of the interlayer tunneling theory is that  $T_c^0$ can be enhanced by the coupling between the
layers. Phrased in the language of the LD model, it means that the parameters of this model
can be changed substantially.  The  phase fluctuations should, however, be similar to those in a
conventional superconductor\cite{Kivelson}. 

\section{$c$-axis conductivity sum rule}
For simplicity, consider a model\cite{Andersen1} in which the microscopic
Hamiltonian  expressing  hopping of electrons along the $c$-axis is
\begin{equation}
H_c=-t_{\perp}\sum_{jl,s}c^{\dagger}_{jl,s}c_{jl+1,s}+{\rm h. c.},
\end{equation}
where the label $j$ refers to the sites of the two-dimensional plane, $l$ refers to the layer
index, and $s$ refers to spin; $c^{\dagger}_{jl,s}$ is the electron creation operator.  In this section, I
focus  only on  single layer materials for which all CuO-planes
are equivalent, such as LSCO, Tl2201, Hg1201, and Bi2201. In the next section, I shall also touch upon multilayer
materials.  

One can now derive  a sum rule\cite{Shastry}. First, the frequency and the
wavevector dependent $c$-axis conductivity can be written as
\begin{equation}
\sigma^c(q_c,\omega,T)=-{1\over Ad}\left({ed\over \hbar}\right)^2 \frac{\langle
-H_c(T)\rangle-\Lambda^c_{\rm ret}(q_c,\omega,T)}{i(\omega+i\delta)},
\end{equation}
where $q_c$ is the momentum transfer perpendicular to the plane, $A$ is the two-dimensional area and $d$
is the separation between the layers. The retarded current-current commutator is 
$
\Lambda^c_{\rm ret}(l,t,T)=-i\theta(t) \langle [j_{H}^c(l,t),j_{H}^c(0,0)]\rangle. 
$
The paramagnetic current operator is defined by
$
j^c(l)=it_{\perp}\sum_{j}(c^{\dagger}_{jl,s}c_{jl+1,s}- {\rm h. c.}),
$
and the corresponding Heisenberg operator, $j^c_{H}$, is defined with respect to the full 
interacting Hamiltonian. The
averages refer to the thermal averages and 
$
\langle H_c(T)\rangle = -t_{\perp}\sum_{j,s}\langle c^{\dagger}_{jl,s}c_{jl+1,s}+{\rm h. c.}\rangle.
$
For optical conductivity, one may set $q_c=0$, and then, noting that the retarded
current-current commutator  is analytic in the upper-half of the complex $\omega$-plane , we arrive at the
$c$-axis conductivity sum rule
\begin{equation}
\int_{-\infty}^{\infty}d\omega\ {\rm Re}\ \sigma^c(\omega,T)={\pi e^2 d^2\over \hbar^2 Ad}\langle
-H_c(T)\rangle,
\label{srule}
\end{equation}
which is a variant of  the well-known $f$-sum rule\cite{Martin}.  Note that it is necessary that the
integral runs between the limits $-\infty$ and $\infty$ to arrive at this sum rule. 

There are a number of noteworthy points.
\begin{itemize}
\item It was argued by Kohn\cite{Kohn} that the
$f$-sum rule does not hold in a metal, because the unbounded position operator is not a valid hermitian
operator. Indeed, all derivations of this sum rule involving the position operator in an extended
system do look suspicious. However, the
$f$-sum rule {\em is}  satisfied\cite{Shastry}. The  reason is that  the $f$-sum rule can be derived by
introducing the exponential operator $e^{i{\bf q\cdot x}}$ and then taking the limit ${\bf q}\to 0$.
Of course, there is no such sum rule if the interaction itself is velocity dependent.
\item  On occasions, this
sum rule is written with finite limits, which is assumed to be some interband gap. This is incorrect\cite{foot3}.
\item The right hand side of the sum rule is the average of the single particle hopping
Hamiltonian. This may be deceptive because it is the true interacting kinetic energy. 
\item The sum rule is satisfied at any temperature $T$.
\item The absence of Galilean invariance on a lattice  allows the charge carrying effective mass to vary with
temperature and interaction. In the continuum limit, such that $d\to 0$, but  $t_{\perp}d^2$ fixed, the
right hand side of Eq.~(\ref{srule}) is ${\pi n e^2\over m}$, where
${\hbar^2\over 2m} = t_{\perp}d^2$, and $n$ is the density of electrons in the planes. In this limit, interactions cannot renormalize the effective mass because the
current operator commutes with the Hamiltonian.   
\end{itemize}

We shall now put this sum rule to good use. For a superconductor, we can write quite generally
\begin{equation}
{\rm Re}\ \sigma^{cs}(\omega,T)=D_c(T)\delta(\omega)+{\rm Re}\ \sigma^{cs}_{\rm reg}(\omega,T).
\end{equation}
The first term signifies the lossless  flow of electrons in the superconducting state, while the second
is the regular (nonsingular) part of the optical conductivity. 
The normal state optical conductivity is nonsingular; so,
the sum rule can be cast into a more useful form:
\begin{eqnarray}
D_c(T)&=&\int_0^{\infty}d\omega \bigg[{\rm Re}\ \sigma^{cn}(\omega,T)-{\rm Re}\ \sigma^{cs}_{\rm
reg}(\omega,T)\bigg]\nonumber\\
&+&{\pi e^2 d^2\over 2 Ad \hbar^2}\bigg[\langle -H_c(T)\rangle_s-\langle -H_c(T)\rangle_n\bigg]
\label{srule2}
\end{eqnarray}

If the $c$-axis kinetic energy is unchanged between the normal and the superconducting
states, as it should be in a conventional layered superconductor, we recover a variant of the
Ferrell-Glover-Tinkham sum rule\cite{Tinkham}.
The missing area between the $c$-axis conductivities of the normal and the superconducting states 
is proportional to the $c$-axis superfluid density.

Frequently, the sum rule in Eq.~(\ref{srule2}) is not   meaningfully applied to high temperature 
superconductors.    
Instead of the true sum rule in Eq.~(\ref{srule2}), the following missing area is considered:
\begin{equation}
D_c'(T)=\int_0^{\infty}d\omega [{\rm Re}\ \sigma^{cn}(\omega,T_c)-{\rm Re}\
\sigma^{cs}_{\rm reg}(\omega,T)].
\label{Dcprime}
\end{equation}
Under what conditions can $D_c'(T)$ be related to the true $c$-axis penetration depth?  It
must be assumed that the $c$-axis kinetic energy must be the same for the normal and the superconducting
states and independent of temperature, with the implicit  assumption that the normal state conductivity
would change very little for all  temperatures $T\le T_c$, if  superconductivity could be suppressed.
For a conventional
superconductor, these assumptions are justified, but
not for cuprates.
First, the change in the $c$-axis  kinetic energy is strikingly evident. Second,
the $c$-axis  resistivity is generically semiconducting and strongly temperature dependent, at least in the
underdoped and optimally doped regimes\cite{Batlogg}. This temperature dependence should persist
if superconductivity could be suppressed, say by applying a magnetic field, and therefore 
the  equality of  the conductivity at $T_c$ and those at  $T\le T_c$ cannot be assumed. For LSCO, this has been
demonstrated experimentally\cite{Ando}.  Even the $ab$-plane resistivity was found to be insulating and
temperature dependent once superconductivity was suppressed by applying strong magnetic fields. These
experiments  bring into question theories that are based on the  assumption that the $T=0$
state is metallic\cite{Graf}. Therefore, we conclude that the
consideration of
$D_c'(T)$ begs the interesting question ``Do electrons change their $c$-axis kinetic energy upon entering
the superconducting state?"

On general grounds, there is  little
we can say about ${\rm Re}\ \sigma^{cn}(\omega,T)$ for $T\le T_c$. 
How do we overcome this impasse? To answer this question, consider the sum rule at zero temperature, 
which can be restated as 
\begin{equation}
D_c(0)\ge {\pi e^2 d^2\over 2 Ad\hbar^2}\left[\langle -H_c(0)\rangle_s-\langle -H_c(0)\rangle_n\right].
\end{equation}
I have assumed that the integral in Eq.~(\ref{srule2}) is positive definite.
This could
be a strict inequality, although I cannot find a rigorous  argument. One can see, however, that at very high
frequencies the two conductivites should approach each other, and, at low frequencies, $\sigma^{cn}_{\rm
reg}(\omega,T=0)\ge \sigma^{cs}_{\rm reg}(\omega,T=0)$, if the superconducting state is at least
partially gapped. 

If the experiments of Ando {\em et al.}\cite{Ando} are taken as an indication, the system, at $T=0$, is
insulating along the $c$-axis. It is plausible, therefore, that  the integral in Eq.~(\ref{srule2}) is
smaller than what one would have guessed for metallic conduction along the $c$-axis. This is because the
frequency dependent $c$-axis conductivity in a non-Fermi liquid is expected to vanish as a power law in
contrast to the Drude behavior.
If this is indeed true, we can make the approximation
\begin{equation}
D_c(0)\approx{\pi e^2 d^2\over 2 Ad\hbar^2}\left[\langle -H_c(0)\rangle_s-\langle
-H_c(0)\rangle_n\right].
\end{equation} 
Defining $n_s^c(0)$ by
$
D_c(0)={\pi n_s^c(0) e^2\over 2m},
$
and $\delta T$ by $\delta T=\left[\langle -H_c(0)\rangle_s-\langle
-H_c(0)\rangle_n\right]$, we get the  simple equation
\begin{equation}
{\hbar^2 n_s^c(0)\over m d^2}\approx{\delta T\over Ad}.
\label{ILT}
\end{equation}
The precise definition of the mass, $m$, is irrelevant because the penetration depth depends only on
$D_c(0)$, that is, only on the combination  $n_s^c(0)e^2\over m$. The left hand side of Eq.~(\ref{ILT}) is of the
order of the confinement kinetic energy of a particle in an one dimensional potential well of width
$d$, consistent with the uncertainty principle.

The $c$-axis penetration depth is given by\cite{foot6} 
\begin{equation}
{1\over\lambda_c^2(0)}={8D_c(0)\over c^2},
\end{equation}
where $c$ is the velocity of light. Therefore, it
satisfies the inequality
\begin{equation}
\lambda_c(0)\le {\hbar c\over e d}{1\over\sqrt{4\pi(\delta T/Ad)}}.
\end{equation}
If we replace $(\delta T/Ad)$ by the condensation energy $U$ of the electrons per unit cell per CuO-layer, including
both spin orientations, and replace the inequality by the equality, we arrive at the approximate expression 
\begin{equation}
\lambda_c(0)\approx {\hbar c\over e d}{1\over\sqrt{4\pi U}}\label{lambdac},
\end{equation}
which is twice as large as that of Anderson\cite{Anderson2}. While Anderson equates the condensation energy to the
Josephson coupling energy, $E_c$, I have  equated it to the change in the kinetic energy.  I believe that this is 
more appropriate because there can be situations in which the condensation energy of the superconductor is not
derived from the change in the kinetic energy, but $E_c$ is finite---conventional Josephson effect, for example.

Anderson\cite{Anderson2} has observed
that  $\lambda_c$ calculated from the procedure outlined above agrees  well with
the measured values. Actually, my expression for $\lambda_c$ in Eq.~(\ref{lambdac}) is a factor of 2 larger, but
this may not be  significant at this time, given the uncertainties involved in extracting the condensation energy
from  the measured specific heat\cite{Loram}.  In LSCO, the $c$-axis reflectivity   exhibits a striking plasma edge
in the superconducting state whose position is readily determined\cite{Uchida}.  As there is little ambiguity in the
measured background dielectric constant, which is approximately 25, the penetration depth can be easily read off
from the plasma edge. In contrast, the analysis based on the missing area, if not properly carried out, will be
flawed\cite{foot5}. For all doping, the agreement found by Anderson is good. It is also reassuring to note that the
penetration depth measured from the plasma edge is  in agreement with the  microwave measurements\cite{Shibauchi}. 
For the single layer Hg1201, the condensation energy is not known from experiments.  Anderson estimated it
from the assumption  that it is proportional to
$T_c^2$. This  yields a  penetration depth in good
agreement with experiments\cite{Cooper}. It must be remembered, however, that this estimate
is subject to a greater uncertainty.  

The  fly in the ointment is the  measurement of Moler {\em et al.}\cite{KAM} 
in the single layer Tl2201. The measured penetration depth is almost a factor of 20 too
large\cite{Anderson2}. Given the  similarities between Tl2201 and Hg1201, this is surprising. However, the
$c$-axis resistivity of Tl2201 is very anomalous; not only does it not show insulating behavior, but it is
linear in its temperature dependence; the magnitude of the resistivity near $T_c$ is enormous, however. In
addition, the material chemistry  of Tl2201 is   quite curious. The optimally doped materials contain
significant interstitial oxygen defects between the two TlO planes, but more surprisingly, they also contain sizable
Cu substitution at the Tl site\cite{Jorgensen}. It may be that there are metallic shorts connecting the CuO planes.
Thus, it is unclear if this measurement reflects the true penetration depth of this material or not.  
The material chemistry of Hg1201 appear to be somewhat different\cite{Jorgensen}. 

It is interesting that the same sum rule  can be turned on its ear to argue that conventional explanations of
$\lambda_c$ are implausible. In Fermi liquid based theories, the change in the $c$-axis kinetic energy must be zero.
The penetration depth is then
\begin{equation}
\lambda_c(0)={c\over \left(8\int_0^{\infty}d\omega [{\rm Re}\ \sigma^{cn}(\omega,0)-{\rm Re}\
\sigma^{cs}_{\rm reg}(\omega,0)]\right)^{1/2}}.
\end{equation}
As argued above, the integral on the right hand side of the denominator is likely to be small. Consequently,  the
penetration depth obtained from this formula is likely to be too large to agree with experiments in LSCO and Hg1201.
Note that this sum rule argument is independent of any microscopic details.

\section{Interlayer enhancement of the mean field transition temperature}
In this section we return to the enhancement of $T_c^0$ in multilayer materials
to compare against the observed systematics, providing further support to the
theory. The inequality derived in the previous section needs only small modifications. The idea is simple.
The coupling between the layers was set by the energy scale     
${\hbar^2 n_s^c(0)\over m d^2}$, but now we have to distinguish between the coupling between the close
layers and the coupling between the distant layers. Let us define them to be $g_{\perp}$ and
$g_{\perp}'$, respectively. Strictly speaking this is a simplification, because the tunneling matrix
elements between the various layers in an unit cell and those in the neighboring cell  are not
all the same.

Imagine that not only the true
transition temperature including fluctuations, but even $T_c^0$ of an
individual layer is very small. I now show that if the coupling between the layers is included, the increase in
$T_c^0$ is negligible in BCS theory. In contrast, the interlayer mechanism leads to a striking enhancement of
$T_c^0$. The mean field equation for single layer materials, due to interlayer coupling, is
\begin{equation}
2g_{\perp}'\chi_{\rm in-plane}(T_c^0)=1,
\label{1layer}
\end{equation}
where $\chi_{\rm in-plane}$ is the in-plane pair susceptibility. For a $n$-layer material, $n\ge 2$, the mean
field equation  is
\begin{equation}
{2(n-1)g_{\perp}+2ng_{\perp}'\over n}\chi_{\rm in-plane}(T_c^0)=1.
\label{2layer}
\end{equation}

We must now determine $\chi_{\rm in-plane}(T)$. Our knowledge of the scale of the coupling energy is
insufficient for this purpose\cite{foot4}. To apply the mean field argument, it is necessary to know the
nature of the coupling between the planes. In particular, it is necessary to know if the Josephson pair
tunneling Hamiltonian is diagonal in the parallel momentum or not. One of the striking aspects of
the interlayer tunneling theory is that it is approximately diagonal in the parallel
momentum\cite{Anderson1,Chakravarty2}. So, the
$\chi_{\rm in-plane}$ to be substituted in these mean field equations must correspond to the momentum for which
this susceptibility is the largest. To see the striking difference caused by this assumption, it is  necessary to
consider only the BCS pair susceptibility. If the coupling Hamiltonian is not diagonal in the parallel momentum, it
is the momentum integrated pair susceptibility that is relevant, which, however,  is only logarithmically
divergent because
\begin{eqnarray}
\chi_{\rm BCS}(T)&=&N(0)\int_0^{\omega_c}{d\varepsilon\over \varepsilon}\tanh({\varepsilon\over 2T}),
\nonumber \\
&=&N(0)\ln(1.14\omega_c/2T).
\end{eqnarray}
where $N(0)$ is the density of states at the Fermi energy and $\omega_c$ is a cutoff of the order of the
Debye energy. When substituted in  Eqs. (\ref{1layer}, \ref{2layer}), the enhancement  of $T_c$ is
negligible because $N(0)g'_{\perp}$, $N(0)g_{\perp}$ are small compared to unity.
In contrast, if the Josephson pair tunneling is diagonal in the parallel momentum, it is the
maximum susceptibility  on the Fermi surface that is relevant. As
$\varepsilon\to 0$ , 
\begin{equation}
\chi_{\rm BCS}(\varepsilon\to 0,T)={1\over 2T},
\end{equation}
diverges faster as $T\to 0$.  When this susceptibility is substituted into Eqs.
(\ref{1layer}) and (\ref{2layer}), it gives rise to far greater enhancements of the transition temperature.  
It can be shown from simple models of a non-Fermi liquid that the
largest in-plane pair susceptibility (not momentum integrated) remains of the same order\cite{Yin},
that is, it is
$a/T$, for T close to $T_c^0$, where
$a$ is a number of order unity. Then, the mean field transition temperatures $T_{c1}^0$, $T_{c2}^0$,
$T_{c3}^0$,
$T_{c4}^0$, for one, two, three, four, \ldots layer materials are given by
$
T_{c1}^0={2g'_{\perp}\over a}$,
$T_{c2}^0=T_{c1}^0+{g_{\perp}\over a}$,
$T_{c3}^0=T_{c1}^0+{4\over 3}{g_{\perp}\over a}$,
$T_{c4}^0=T_{c1}^0+{3\over 2}{g_{\perp}\over a}$,
etc; the sequence $1, 4/3, 3/2, \ldots$ converges to 2. This pattern of the systematic enhancement of the transition
temperatures  of the multilayer materials are in accord with experiments.
\section{Conclusion}
The purpose of this paper has been to present some very general arguments in favor of the interlayer
tunneling theory. There are two significant outcomes of the present paper. The first  concerns the
nature of the 2$D$ to 3$D$ crossover in the superconducting state. It was shown that the interlayer
tunneling mechanism can enhance the microscopic superfluid stiffness in a way that is not possible in 
conventional theories. This stiffness sets the scale at which the amplitude of the order parameter
forms. On general grounds, it is  difficult to settle whether or not phase fluctuations are important;
experimental evidence  on this question appears to be mixed.  If, however,  they are assumed to be
present, as has been argued by Emery and Kivelson\cite{Kivelson}, they can be included by combining
interlayer tunneling theory with the Lawrence-Doniach model. The pseudogap observed in the
underdoped  materials in that case would be the superconducting gap calculated within the interlayer
tunneling theory, while the true $T_c$ will be  determined by the phase fluctuations. 

The second outcome of the present paper is the resolution of the paradoxical
interpretations  of the
$c$-axis optical measurements, universally evident in the literature. On  the one hand, it appeared that
one could obtain the correct estimates of the
$c$-axis penetration depths only from the change in the kinetic energy of the electrons as they enter
the superconducting state\cite{Anderson2}, on the other hand, the same results were apparently obtained from
the
$c$-axis conductivity sum rule ignoring the change in the kinetic energy\cite{Uchida}. The resolution is
that, until now, the sum rule has not been meaningfully applied to high temperature superconductors. The
paradox disappears with the correct interpretation of the sum rule. The evidence for the change in
electron's kinetic energy, an essential element of the interlayer tunneling theory, appears to be strong in
LSCO, reasonably convincing in Hg1201, and  nonexistent in
Tl2201 on the basis of the recent measurements\cite{KAM}. In regard to Tl2201, important materials 
questions remain.

Future measurements of the $c$-axis
optical conductivity and the penetration depth in both single and multilayer materials will be
valuable. In particular, I would like to suggest that these experiments be carried out on the single
layer Bi2201, which in many respects is as anomalous as its high-$T_c$ cousins. If possible, the
optical measurements should be carried out in the presence of a magnetic field necessary to suppress
superconductivity. In this low $T_c$ material, the required magnetic field should be considerably
smaller than in the experiments of Ando {\em et al.}\cite{Ando} Moreover, the normal state can be
pursued and measured more precisely to lower temperatures.

\section{Acknowledgement}
I thank P. W. Anderson,  D. J. Scalapino, and especially D. Basov, S. Kivelson, and
K. A. Moler for discussions. This work was supported by a grant from the
National Science Foundation: DMR-9531575.


\begin{references}
\bibitem{foot1}This is only true for higher energy processes. For lower energy processes, there 
is incoherent hopping
of electrons between the planes that gives rise to zero frequency  conductivity 
at finite temperatures.
\bibitem{foot2}The phase across a large area, large capacitance, Josephson junction 
is a macroscopic variable obeying classical dynamical equation with the Hamiltonian determined from the
coupling energy. For small capacitance junctions, quantum fluctuations of the phase become
important (See Ref.\cite{Tinkham} and references therein.)
\bibitem{foot3}The hopping Hamiltonian is an effective model, where all the effects of higher energy bands are in
principle downfolded by a method akin to a renormalization group procedure\cite{Andersen1}. This implies that this
effective Hamiltonian is applicable only to low energy processes below a cutoff presumably of the order of the
interband transition.  But once this model is determined, theoretically there are no more degrees of freedom left
and mathematically the integral in Eq.~(\ref{srule}) has to be between $-\infty$ and $\infty$ for the sum rule to be
correct. This means that the conductivity $\sigma^c(\omega)$ calculated from this effective Hamiltonian cannot be
the physical conductivity for scales of the order or larger than the high energy cutoff implied in the effective
Hamiltonian. This makes it problematic to accurately estimate the missing area as only the order of magnitude of the
cutoff is known.
\bibitem{foot6}It is useful to
elaborate on this further. From the singular part of the real part of the conductivity, we conclude from the
Kramers-Kronig relation that 
$
\omega{\rm Im}\ \sigma(\omega)={2D_c(T)\over \pi}
$.
Consider now the Josephson expression for the current density in its gauge invariant form, which is
$
J=J_c\sin(\delta\theta-{2ed{\cal A}\over \hbar c})
$,
assuming a uniform vector potential $\cal A$ between the two layers separated by $d$; $J_c$ is the critical current
density. If we now set
$\delta\theta=0$ and take the limit ${\cal A}\to 0$, then we get
$
J=-{2J_ced \over \hbar c}{\cal A}
$.
Equating this to the defining equation for  the penetration depth,
$
J=-{c\over 4\pi\lambda_c^2}{\cal A},
$
we get
$
\lambda_c=\left({\Phi_0c\over 8\pi^2J_cd}\right)^{1/2},
$
which is a frequently used expression for layered superconductors described by the LD model (See, for example, Moler
{\em et al.}\cite{KAM} and references therein.). The remaining task is to relate
$J_c$ to the imaginary part of the conductivity. This is easily done by choosing the vector potential to be time
dependent, that is, ${\cal A}(t)={\cal A}_{\omega}e^{-i\omega t}$. Then, following the same line of reasoning as
above, we get
$
\omega\ {\rm Im}\sigma(\omega)={2edJ_c\over \hbar}.
$
The penetration depth can now be expressed as
$
\lambda_c={c\over \sqrt{4\pi\omega {\rm Im}\sigma(\omega)}}
$.
We can now substitute the expression for the imaginary part of the conductivity to find the desired expression for
the penetration depth: $\lambda_c=c/(8D_c(T))^{1/2}$.
\bibitem{foot4}This is also true for classical statistical mechanical problems discussed in Sec. 2.
\end{references}
\end{document}